\documentclass[twocolumn,floatfix,aps,prb]{revtex4-2}
\usepackage{graphicx}
\usepackage{amsmath}
\usepackage{amssymb}
\usepackage{bm}
\usepackage{hyperref}

\begin{document}

\title{Casimir-Josephson force on a point contact between two superconductors}
\author{C. W. J. Beenakker}
\affiliation{Instituut-Lorentz, Universiteit Leiden, P.O. Box 9506, 2300 RA Leiden, The Netherlands}
\date{October 2023}

\begin{abstract}
We calculate the elongation or contraction force $F$ on a point contact (length $L$) connecting two superconductors with a phase difference $\phi$. When $L$ is small compared to the superconducting coherence length $\xi_0$ this force is given by $F=-(\Delta_0/\pi\xi_0)\ln(\xi_0/L)\cos\phi$ per spin-degenerate transverse mode. Quantum fluctuations in states from the continuous spectrum outside the superconducting gap $\Delta_0$ give the dominant contribution to this force, which may be understood as the superconducting counterpart of the electromagnetic Casimir force. We compare with earlier work that only included contributions from the discrete spectrum of Andreev levels.
\end{abstract}
\maketitle

\section{Introduction}

The free energy $\Omega$ of a Josephson junction, a weak link between two superconductors, depends on the superconducting phase difference $\phi$ across the junction as well as on the junction length $L$. The derivative of $\Omega$ with respect to $\phi$ gives the supercurrent $I=(2e/\hbar)d\Omega/d\phi$, the derivative with respect to $L$ gives the elongation or contraction force $F=-d\Omega/dL$ on the junction \cite{Kri04a,Kri04b,Par12,Bul05,Wir06}. 

This force is the superconducting counterpart of the electromagnetic Casimir force \cite{Cas48}, as one can see from the expression for a single-mode junction in the large-$L$ limit \cite{Kri04a,Cau02,Fuc07,Zha08},
\begin{equation}
F=-\frac{g\hbar v_{\rm F}}{24\pi L^2}(\pi^2-3\phi^2),\;\;|\phi|<\pi,\label{Flongjunction}
\end{equation}
with $v_{\rm F}$ the Fermi velocity and $g$ a factor that counts spin and other degeneracies.
For $\phi=0$, $g=1$, $v_{\rm F}=c$ this is precisely the attractive Casimir force produced by vacuum fluctuations of a scalar one-dimensional wave between perfectly reflecting metallic mirrors \cite{Plu86,Mos97,Boy03,note1}. In the Josephson junction the superconducting condensate at zero temperature plays the role of the electromagnetic vacuum.

\begin{figure}[tb]
\centerline{\includegraphics[width=0.9\linewidth]{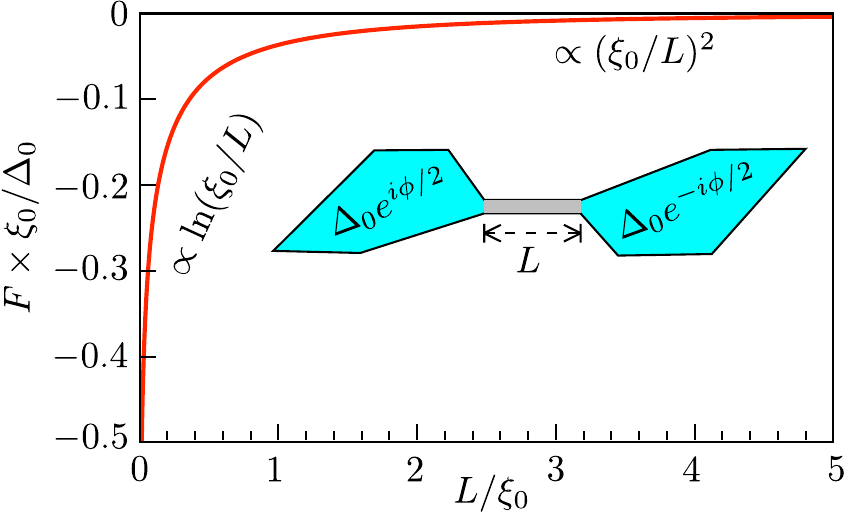}}
\caption{Dependence of the force on a  Josephson junction on the junction length $L$ (computed from Eq.\ \eqref{omegaintegral} for superconducting phase difference $\phi=0$). The force is negative, so it is a contraction force, crossing over to an elongation force for $\phi\gtrsim\pi/2$. The plot refers to a single transverse mode. The inverse-square large-$L$ asymptotic is from Ref.\ \onlinecite{Kri04a}, the logarithmic small-$L$ asymptotic is obtained here.
}
\label{fig_layout}
\end{figure}

The result \eqref{Flongjunction} holds if the junction is long compared to the superconducting coherence length $\xi_0=\hbar v_{\rm F}/\Delta_0$ (for a superconducting gap $\Delta_0$ and assuming ballistic transport through the junction). The short-junction regime $L\ll\xi_0$ has no electromagnetic analogue. This regime is relevant for a point contact Josephson junction (as in Fig.\ \ref{fig_layout}). In what follows we will study that regime theoretically.

One might surmise that the point contact length $L$ should be replaced by an effective length $L+\xi_0$, to estimate $F\simeq \hbar v_{\rm F}/\xi_0^2$ in the short-junction regime. The pioneering work by Krive \textit{et al.} \cite{Kri04a} arrived at this expression. As we shall see, a more complete calculation introduces a logarithmic $L$-dependence $\propto\ln(\xi_0/L)$, due to $\phi$-dependent contributions from the continuous spectrum (neglected in Ref.\ \onlinecite{Kri04a}).

The insight that the continuous spectrum gives a non-analytic contribution $\propto\ln L$ to the Josephson free energy is not new \cite{Asl69,Bro97,Lev06}. What is special about the Casimir-Josephson force is that the $\ln L$ term is the \textit{leading contribution} in the short-junction limit --- while for the supercurrent the leading contribution is $L$-independent.

The outline of this paper is as follows. In the next section we summarize the scattering matrix representation \cite{Bro97,Bee91} of the Josephson free energy $\Omega$, which is a convenient starting point because it treats the discrete and continuous spectrum on the same footing. We apply this to a ballistic point contact in Sec.\ \ref{sec_ballistic}, to calculate the force $F=-d\Omega/dL$ for $L\ll\xi_0$. The effect of a tunnel barrier (transmission probability $\Gamma$) in the point contact is then treated in Sec.\ \ref{sec_tunneling}. The results of these two sections differ from those of Ref.\ \onlinecite{Kri04a} in two aspects, both attributable to the continuous spectrum: an enhancement of the force by a factor $\ln(\xi_0/L)$ and a reduction of the force by a factor $\sqrt\Gamma$.

\section{Scattering formulation}

The free energy of a Josephson junction at temperature $T$ is given by \cite{Bar69}
\begin{equation}
\Omega=-gT\int_0^\infty d\varepsilon\,\rho(\varepsilon)\ln\bigl[2\cosh(\varepsilon/2T)\bigr].\label{omegadef}
\end{equation}
(We set $\hbar$ and Boltzmann's constant $k_{\rm B}$ equal to unity.) The density of states $\rho(\varepsilon)$ refers to the electron-hole symmetric spectrum of the Bogoliubov-De Gennes Hamiltonian $H_{\rm BdG}$. We ignore $\phi$-independent contributions to the free energy from the superconducting reservoirs \cite{Bee92}.

The spectrum of $H_{\rm BdG}$ consists of bound states (Andreev levels) for $\varepsilon<\Delta_0$ and a continuous spectrum for $\varepsilon>\Delta_0$. Scattering theory \cite{Bee91} includes both contributions in the determinantal formula
\begin{equation}
\rho(\varepsilon)=-\frac{1}{\pi}\operatorname{Im}\frac{d}{d\varepsilon}\ln\operatorname{det}\bigl[1-R_{\rm A}(\varepsilon+i0^+)S_{\rm N}(\varepsilon+i0^+)\bigr],\label{rhodef}
\end{equation}
in terms of the Andreev reflection matrix $R_{\rm A}$ and the scattering matrix $S_{\rm N}$ of the junction in the normal state.

The key physical ingredient in this representation of the density of states is the separation of length scales between Andreev reflection \cite{And64}, the conversion of an electron into a hole on the length scale of the coherence length $\xi_0$, and normal scattering processes on the length scale of the Fermi wave length $\lambda_{\rm F}$. The ratio $\lambda_{\rm F}/\xi_0\simeq\Delta_0/E_{\rm F}\ll 1$ in a superconductor.

The Andreev reflection matrix has the block structure
\begin{equation}
\begin{split}
&R_{\rm A}(\varepsilon)=\alpha(\varepsilon)\begin{pmatrix}
0&r_{\rm A}\\
r^\ast_{\rm A}&0
\end{pmatrix},\;\;
r_{\rm A}=
\begin{pmatrix}
e^{i\phi/2}&0\\
0&e^{-i\phi/2}
\end{pmatrix},\\
&\alpha(\varepsilon)=\varepsilon/\Delta_0-i\sqrt{1-\varepsilon^2/\Delta_0^2}.
\end{split}\label{RAdef}
\end{equation}
The blocks $r^\ast_{\rm A}$ and $r_{\rm A}$ describe Andreev reflection from electron to hole and from hole to electron, respectively. The diagonal elements of the submatrix $r_{\rm A}$ refer to Andreev reflection from the left and right superconductor, at a phase $\pm\phi/2$.
Andreev reflection happens with unit probability for $\varepsilon<\Delta_0$, at larger energies $\alpha(\varepsilon)=\varepsilon/\Delta_0-\sqrt{\varepsilon^2/\Delta_0^2-1}$ decays to zero. The normal scattering matrix $S_{\rm N}$ is block-diagonal,
\begin{equation}
S_{\rm N}(\varepsilon)=\begin{pmatrix}
s_0(\varepsilon)&0\\
0&s_0^\ast(-\varepsilon)
\end{pmatrix},\label{SNdef}
\end{equation}
with electron scattering matrix $s_0$ unitary at all energies. 

Substitution of Eqs.\ \eqref{RAdef} and \eqref{SNdef} into Eq.\ \eqref{rhodef} gives the determinant \cite{Bee91}
\begin{equation}
\begin{split}
&\rho(\varepsilon)=-\frac{1}{\pi}\operatorname{Im}\frac{d}{d\varepsilon}\ln\operatorname{det}\bigl[1-M(\varepsilon+i0^+)\bigr],\\
&M(\varepsilon)=\alpha(\varepsilon)^2 r_{\rm A}^\ast s_0(\varepsilon) r_{\rm A}^{\vphantom{\ast}} s_0^\ast(-\varepsilon)].
\end{split}
\end{equation}

As a final step, the slowly converging integration over energies in Eq.\ \eqref{omegadef} can be transformed into a more rapidly converging sum over Matsubara frequencies. The resulting free energy is given by \cite{Bro97} 
\begin{equation}
\Omega=-gT\sum_{p=0}^\infty\ln\det[1-M(i\omega_p)],\;\;\omega_p=(2p+1)\pi T.\label{omegap}
\end{equation}
In the zero-temperature limit the sum over $p$ can be replaced by an integral over $\omega$,
\begin{equation}
\lim_{T\rightarrow 0}\Omega=-\frac{g}{2\pi}\int_0^\infty d\omega\,\ln\det[1-M(i\omega)].\label{OmegaMatsubara}
\end{equation}

\section{Ballistic point contact}
\label{sec_ballistic}

For simplicity we consider an $N$-mode point contact which does not mix the transverse modes. For each mode the scattering matrix $s_0$ is a $2\times 2$ matrix of reflection amplitudes (on the diagonal) and transmission amplitudes (off-diagonal). 

In this section we assume ballistic transport, so only the transmission amplitudes are nonzero,
\begin{equation}
s_0(\varepsilon)=e^{ik(\varepsilon)L}\begin{pmatrix}
0&1\\
1&0
\end{pmatrix}.\label{s0def}
\end{equation}
Here $k(\varepsilon)$ is the momentum of the $n$-th electron mode in the point contact region between the superconductors. For $\varepsilon\ll E_{\rm F}$ we may linearize its energy dependence, $k(\varepsilon)=k_{\rm F}+\varepsilon/v_{\rm F}$, with $k_{\rm F}$ and $v_{\rm F}$ the momentum and velocity of the mode at the Fermi level.

The contribution of each mode to the free energy \eqref{OmegaMatsubara} then takes the form
\begin{align}
&\Omega=-\frac{g\Delta_0}{2\pi}\int_0^\infty d\omega\,\ln\bigl[1+\beta(\omega)^4 e^{-4 \omega L/\xi_0}\nonumber\\
&\qquad\qquad\qquad+2 \beta(\omega)^2 e^{-2 \omega L/\xi_0} \cos \phi\bigr]\nonumber\\
&\;\;=-\frac{g\Delta_0}{\pi}\operatorname{Re}\int_0^\infty d\omega\,\ln\bigl[1+\beta(\omega)^2 e^{i\phi-2 \omega L/\xi_0}\bigr],\label{omegaintegral}\\
&\beta(\omega)=\omega-\sqrt{1+\omega^2}.
\end{align}
We seek the force $F=-d\Omega/dL$.

For $L\gg\xi_0$ we may set the function $\beta(\omega)$ to unity in Eq.\ \eqref{omegaintegral}, and we recover the known result \eqref{Flongjunction} for the force in the long-junction regime, since
\begin{equation}
\operatorname{Re}\int_0^\infty dx\,\ln\bigl(1+ e^{i\phi-x}\bigr)=\frac{\pi^2}{12}-\frac{\phi^2}{4},\;\;|\phi|<\pi.
\end{equation}

In the opposite short-junction regime $L\ll\xi_0$ the $L$-dependence of the integral \eqref{omegaintegral} is governed by the large-$\omega$ range, where the integrand decays as $\omega^{-2}e^{-2\omega L/\xi_0}$. This gives a contribution to the force integral which decays as $\omega^{-1}e^{-2\omega L/\xi_0}$, producing a logarithmic $L$-dependence upon integration over $\omega$,
\begin{equation}
F=-\frac{g\Delta_0}{2\pi\xi_0}\bigl[\ln(\xi_0/L)\cos\phi+{\cal O}(1)],\;\;L\ll\xi_0.\label{Flog}
\end{equation}
A calculation to higher order in $L/\xi_0$ in the Appendix gives
\begin{align}
F={}&-\frac{g\Delta_0}{2\pi\xi_0}\bigl[\ln(\xi_0/L)\cos\phi-\gamma_{\rm Euler}\cos\phi\nonumber\\
&-\tfrac{1}{2}(1-\phi\sin\phi)+{\cal O}(L)\bigr],\;\;|\phi|<\pi.\label{Fexpansionresult}
\end{align}

\begin{figure}[tb]
\centerline{\includegraphics[width=0.8\linewidth]{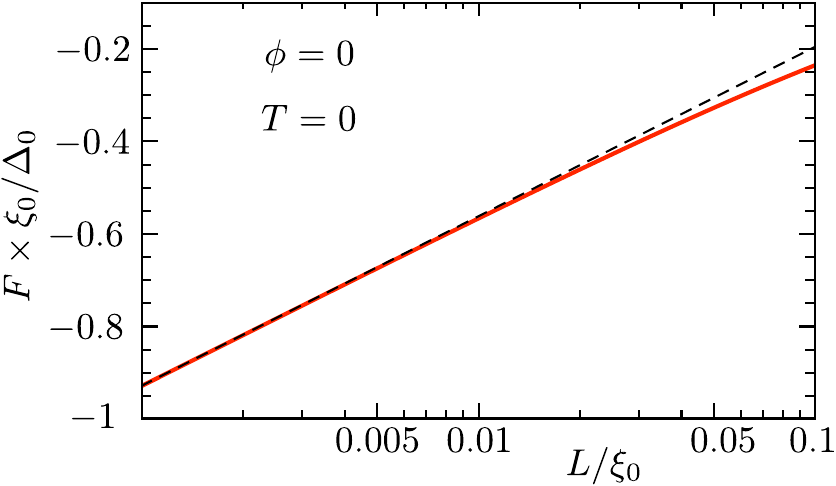}}
\centerline{\includegraphics[width=0.8\linewidth]{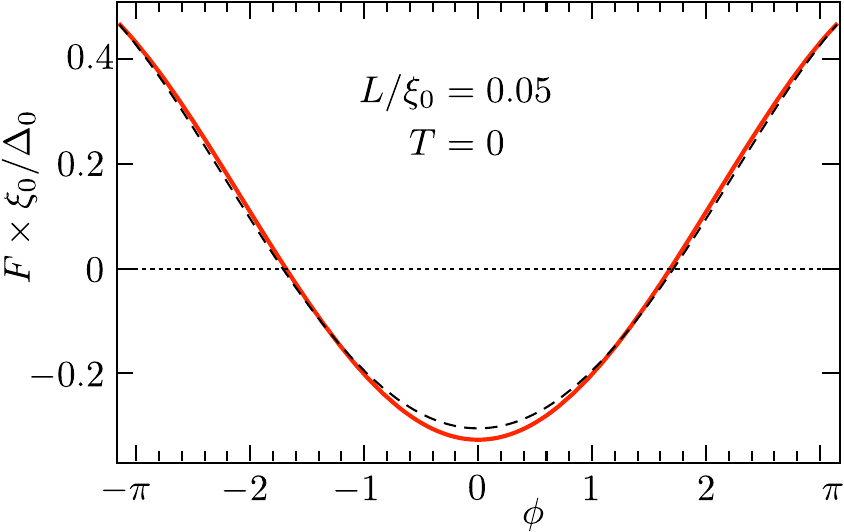}}
\caption{Dependence of the force on a Josephson junction on the junction length $L$ for $\phi=0$ (top plot, on a log-linear scale) and dependence on the phase difference $\phi$ for $L/\xi_0=0.05$ (bottom plot). The red solid curves are computed from Eq.\ \eqref{omegaintegral}, the black dashed curves are the small-$L$ asymptotic \eqref{Fexpansionresult}. The values of the force refer to a single transverse mode. The $\phi$-dependent plot repeats with $2\pi$-periodicity outside of the interval $-\pi<\phi<\pi$. These are results at zero temperature.
}
\label{fig_plots}
\end{figure}

\begin{figure}[tb]
\centerline{\includegraphics[width=0.8\linewidth]{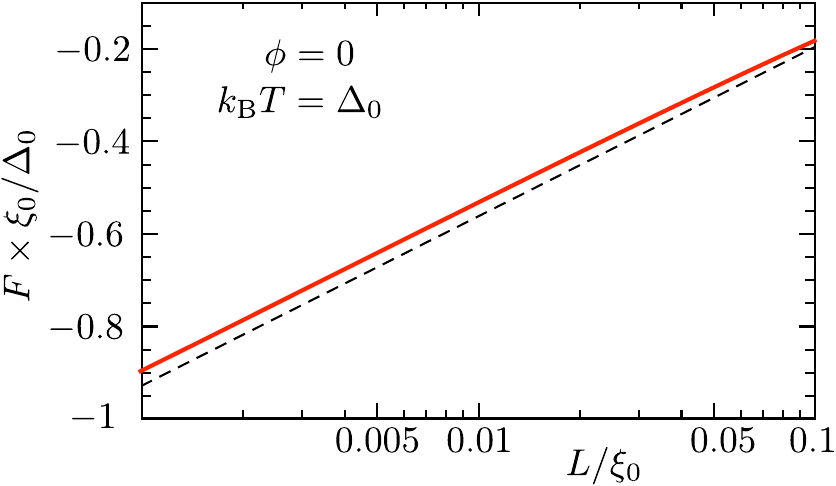}}
\centerline{\includegraphics[width=0.8\linewidth]{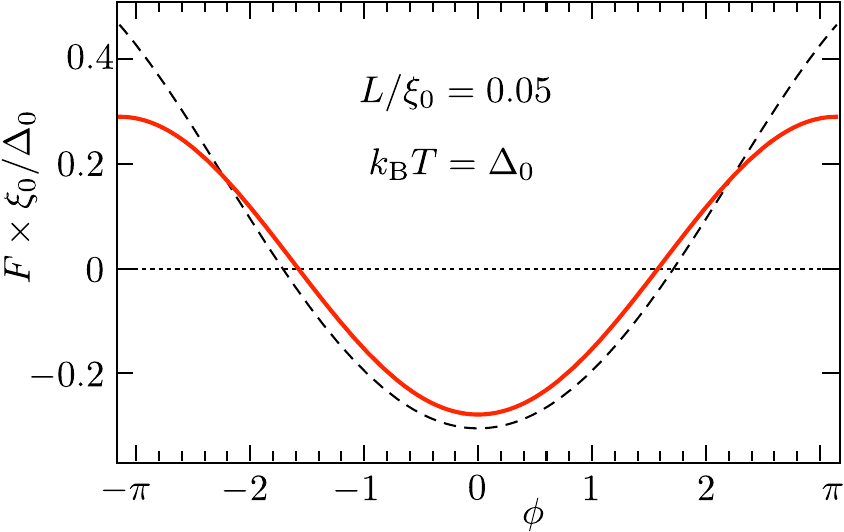}}
\caption{Same as Fig.\ \ref{fig_plots}, but now at temperature $T=\Delta_0$, when the integral over $\omega$ is replaced by the sum over $\omega_p=(2p+1)\pi T$ in Eq.\ \eqref{omegap}. The dashed curves are the zero-temperature result \eqref{Fexpansionresult}, included for comparison.
}
\label{fig_plotsb}
\end{figure}

In Fig.\ \ref{fig_plots} we compare the full expression \eqref{omegaintegral} with the small-$L$ asymptotics \eqref{Fexpansionresult} for the Casimir-Josephson force. This is at zero temperature. The effect of a nonzero temperature $T\lesssim\Delta_0$ is shown in Fig.\ \ref{fig_plotsb}: It is predominantly a rounding of the cusp in the $\phi$-dependence at $\phi=\pm\pi$. The scaling with $L$ is not significantly affected.

The logarithmic $L$-dependence of the force is due to the contribution from the continuous spectrum, the energy range $\varepsilon>\Delta_0$ in the free energy integral \eqref{omegadef}. To see this, note that the discrete spectrum consists for $L\ll\xi_0$ of a single excitation energy $\varepsilon_0>0$ per transverse mode, at \cite{Kri04a}
\begin{equation}
\varepsilon_0=\Delta_0\left(\cos(\phi/2)-\frac{L}{2\xi_0}|\sin\phi|+{\cal O}(L^2)\right),\;\;|\phi|<\pi.\label{eps0def}
\end{equation}
This contributes to the Casimir-Josephson force an $L$-independent amount $d\varepsilon_0/dL\simeq \Delta_0/\xi_0$, without the $\ln (\xi_0/L)$ factor.

The integral over the continuous spectrum is cut off by the band width $E_{\rm F}$. This cut off is ineffective if $L>\lambda_{\rm F}$, but for smaller junction lengths the $\ln(\xi_0/L)$ increase saturates at $\ln(\xi_0/\lambda_{\rm F})\propto \ln(E_{\rm F}/\Delta_0)$.

A final remark for this section: As one can see by comparing Eqs.\ \eqref{Flog} and \eqref{eps0def}, the continuous spectrum changes the $\phi$-dependence of the force from $|\sin\phi|$ to $\cos\phi$. The removal of the cusp singularity at $\phi=0$ changes how the force responds to the presence of a tunnel barrier in the point contact, as we shall see in the next section.

\section{Point contact with a tunnel barrier}
\label{sec_tunneling}

A tunnel barrier with transmission probability $\Gamma\in(0,1)$ at the center of the point contact modifies the scattering matrix \eqref{s0def} and zero-temperature free energy \eqref{OmegaMatsubara} as follows:
\begin{align}
&s_0(\varepsilon)=e^{i(k_{\rm F}+\varepsilon/v_{\rm F})L}\begin{pmatrix}
\sqrt{1-\Gamma}&\sqrt{\Gamma}\\
\sqrt{\Gamma}&-\sqrt{1-\Gamma}
\end{pmatrix},\\
&\Omega=-\frac{g\Delta_0}{2\pi}\int_0^\infty d\omega\,\ln\bigl[1+\beta(\omega)^4 e^{-4 \omega L/\xi_0}\nonumber\\
&\qquad\qquad+2 \beta(\omega)^2 e^{-2 \omega L/\xi_0} \bigl(1-2\Gamma\sin^2(\phi/2)\bigr)\bigr].
\end{align}
So the expressions from the previous section can be used with the substitution
\begin{equation}
\cos\phi\mapsto 1-2\Gamma\sin^2(\phi/2).\label{substitution}
\end{equation}

For $\Gamma\ll 1$ the $\phi$-dependence of the force on the Josephson junction is reduced by a factor $\Gamma$. In particular, in the short-junction regime $L\ll\xi_0$ one has, in view of Eq.\ \eqref{Flog},
\begin{equation}
F=\frac{g\Delta_0}{\pi\xi_0}\ln(\xi_0/L)\Gamma\sin^2(\phi/2)+\phi\text{-independent terms}.\label{Floggamma}
\end{equation}

Notice that the same substitution \eqref{substitution} would transform $|\sin\phi|$ into $2\sqrt{\Gamma}|\sin(\phi/2)|$ for $\Gamma\ll 1$. This explains why Ref.\ \onlinecite{Kri04a} found that a tunnel barrier reduces the $\phi$-dependence of the force by a factor $\sqrt{\Gamma}$, rather than by a factor $\Gamma$.

\section{Conclusion}
\label{sec_conclusion}

In summary, we have calculated the Casimir-like force on a point contact Josephson junction, focusing on the regime that the length $L$ of the junction is short compared to the superconducting coherence length $\xi_0$. In the limit $L/\xi_0\rightarrow 0$ the supercurrent through the junction is fully determined by the discrete spectrum of Andreev levels, confined to the junction by the superconducting gap $\Delta_0$. 

We have shown that the same does not apply to the elongation or contraction force on the junction: the continuous spectrum of states outside the gap qualitatively modifies the dependence of the force on both the junction length $L$ and the superconducting phase difference $\phi$: The $L$-dependence acquires a logarithmic factor $\ln(\xi_0/L)$ and the $\phi$-dependence becomes more sensitive to the presence of a tunnel barrier in the point contact (decreasing $\propto\Gamma$ rather than $\propto\sqrt{\Gamma}$ with the tunnel probability $\Gamma$).

In a ballistic point contact ($\Gamma=1$) with $N$ spin-degenerate transverse modes ($g=2$) the difference $\delta F$ between the Casimir-Josephson force at $\phi=0$ and $\phi=\pi$ is given for $L\ll\xi_0$ by
\begin{equation}
\delta F=\frac{2N\Delta_0}{\pi\xi_0}\ln(\xi_0/L).
\end{equation}
The logarithmic enhancement factor saturates at $\ln(E_{\rm F}/\Delta_0)$ for $L\lesssim\lambda_{\rm F}$. This will be at best a factor of 10, and since $\Delta_0/\xi_0\lesssim 10^{-15}\,\rm{N}$ one needs a large number of modes, $N\gtrsim 10^4$, to reach a measurable force in the 0.1~nN range.

An additional complication, pointed out by Krive \textit{et al.} \cite{Kri04a}, is that $N$ may itself be dependent on the length of the junction (decreasing with increasing $L$ if the junction is elongated at constant volume). This will induce a $\phi$-dependent contraction force $\simeq (dN/dL)\Delta_0$ of a more mundane origin than the Casimir-Josephson force originating from the $L$-dependence of the Andreev spectrum. A coupling between $N$ and $L$ would thus need to be avoided. Other elastic forces, e.g. due to a bending of the point contact, are $\phi$-independent and therefore do not contribute to $\delta F$, but a succesful measurement of the Casimir-Josephson force remains challenging.

In closing we remark that our result \eqref{Fexpansionresult} also implies a non-analytic finite-$L$ correction to the supercurrent of a ballistic point contact,
\begin{align}
&I=\frac{2e}{\hbar}\frac{g\Delta_0}{4}\sin(\phi/2)\nonumber\\
&\;-\frac{2e}{\hbar}\frac{g\Delta_0}{2\pi}\frac{L}{\xi_0}\bigl(\ln (\xi_0/L) \sin \phi-\tfrac{1}{2}\phi \cos \phi\bigr),\;\;|\phi|<\pi,
\end{align}
which complements the known result for a diffusive point contact \cite{Lev06,note2}.

\acknowledgments

This project has received funding from the European Research Council (ERC) under the European Union's Horizon 2020 research and innovation programme. I have benefited from discussions with A. R. Akhmerov, I. Araya, and T. Vakhtel.

\appendix

\section{Short-junction limit of the force integral}

In view of Eq.\ \eqref{omegaintegral}, the Casimir-Josephson force $F=-d\Omega/dL$ on the Josephson junction at zero temperature is given by the integral
\begin{equation}
\begin{split}
&F=-\frac{2g\Delta_0}{\pi\xi_0}\operatorname{Re}f(L/\xi_0),\\
&f(L)=\int_0^\infty d\omega\,\frac{\omega\beta(\omega)^2e^{i\phi-2 \omega L}}{1+\beta(\omega)^2 e^{i\phi-2 \omega L}}.
\end{split}\label{Ffdef}
\end{equation}
Note that $\omega\beta(\omega)^2\rightarrow \tfrac{1}{4}\omega^{-1}$ for $\omega\rightarrow\infty$, so a logarithmic singularity prevents us from directly taking  the short-junction limit $L/\xi_0\rightarrow 0$.

We isolate the logarithmic singularity by subtracting
\begin{align}
h(L)={}&\int_0^\infty d\omega\,(1+4\omega)^{-1}e^{i\phi-2 \omega L}\nonumber\\
&=-\tfrac{1}{4} \operatorname{Ei}(-L/2) e^{L/2+i \phi}.
\end{align}
The difference has the short-junction limit (for $|\phi|<\pi$)
\begin{equation}
\lim_{L\rightarrow 0}\operatorname{Re}\bigl[f(L)-h(L)\bigr]=-\tfrac{1}{8} (1+(2\ln 2)\cos\phi-\phi \sin \phi).
\end{equation}
Adding the small-$L$ expansion of the exponential integral function $\operatorname{Ei}(z)=-\int_{-z}^\infty t^{-1}e^{-t}\,dt$,
\begin{equation}
\operatorname{Re}h(L)=-\tfrac{1}{4}\bigl[\ln(L/2)+\gamma_{\rm{Euler}}\bigr]\cos\phi+{\cal O}(L),
\end{equation}
we arrive at
\begin{equation}
\operatorname{Re}f(L)=-\tfrac{1}{4}(\ln L+\gamma_{\rm Euler})\cos\phi-\tfrac{1}{8}(1-\phi\sin\phi)+{\cal O}(L).\label{fLexpansion}
\end{equation}

Substitution of the expansion \eqref{fLexpansion} into Eq.\ \eqref{Ffdef} gives the result \eqref{Fexpansionresult} from the main text.

Eq.\ \eqref{fLexpansion} should be repeated with $2\pi$-periodicity outside of the interval $-\pi<\phi<\pi$. The $L$-independent $\phi\sin\phi$ term then produces a cusp singularity at $\phi=\pm\pi$ while the term $\propto\ln L$ is smooth at $\phi=\pm\pi$. This can be understood because the cusp is due to states in the gap crossing the Fermi level, while the $\ln L$ factor is due to states outside of the gap, so the two types of singularities do not coexist in the same term.


\begin{thebibliography}{99}
\bibitem{note1} The $\phi$-dependence of the Casimir-Josephson force \eqref{Flongjunction} can be checked with the corresponding expression for the supercurrent, $dF/d\phi=-(\hbar/2e)dI/dL$, in agreement with the known result \cite{Bar72} $I=(gev_{\rm F}/L)\phi/2\pi$ in the long-junction limit.
\bibitem{note2} In a diffusive point contact (mean free path $\ell\ll L$) the force is dominated by the $L$-dependence of the conductance: $F\simeq -\Delta_0(h/e^2)dG/dL\simeq N\Delta_0 \ell/L^2$.
\end{thebibliography}
\end{document}